\documentclass[twocolumn,prl,showpacs]{revtex4}
\usepackage{psfrag}
\usepackage{graphicx}
\usepackage{amsmath}
\usepackage{amssymb}
\usepackage{times}

\begin{document}
\title{Impact of Exchange-Correlation Effects on the $IV$
  Characteristics of a Molecular Junction}
\author{K. S. Thygesen}
\affiliation{Center for Atomic-scale Materials Design (CAMD),
Department of Physics, Technical University of Denmark, DK - 2800 Kgs.
Lyngby, Denmark}

\date{\today}

\begin{abstract}
  The role of exchange-correlation effects in non-equilibrium
  quantum transport through molecular junctions is assessed by
  analyzing the $IV$ curve of a generic two-level model using self-consistent
  many-body perturbation theory (second Born and $GW$ approximations) on the Keldysh contour. For weak
  molecule-lead coupling we identify a mechanism
  which can lead to anomalously strong peaks in the $dI/dV$ due to a
  bias-induced interplay between the position of the HOMO and LUMO
  levels. The effect is suppressed by self-interaction errors and is
  therefore unlikely to be observed in standard transport calculations
based on density functional theory. Inclusion of dynamic correlations lead to
substantial renormalization of the energy levels. In particular, we
find a strong enhancement of quasi-particle (QP) scattering at finite
bias which reduces the QP
lifetimes significantly with a large impact on the $IV$ curve. 
\end{abstract}

 \maketitle Over the last decade it has
become possible to
contact single molecules by metallic electrodes and measure the $IV$
characteristic of the resulting nano
junction~\cite{reed97,smit02,reichert_weber02}. Experiments of this
kind can be seen as the first step towards the realization of a
molecule based electronics. On a more fundamental level, the $IV$
characteristics provide a spectroscopic fingerprint of the
molecular junction containing information about the positions and lifetimes
of the electronic energy levels. In view of this, the interpretation of
$IV$ curves in
terms of the electronic structure of the junction represents a fundamental
challenge for molecular electronics.

So far, almost all \emph{ab intio} calculations of conductance in molecular
junctions have been based on the single-particle Kohn-Sham (KS) scheme of
Density Functional Theory (DFT)~\cite{brandbyge02,thygesen_bollinger,nitzan03}. This
approach has been successfully applied to junctions characterized by
strong coupling between molecule and
leads~\cite{prl_thygesen_h2,prl_pt_osc}, but has generally been found
to overestimate the low-bias conductance of larger and more weakly coupled
molecules~\cite{neaton,tao_nanolett04,stokbro03}.  Recently, this
shortcoming of the DFT approach has been attributed to the presence of self-interaction (SI)
errors in the standard exchange-correlation (xc) functionals~\cite{burke,toher07}.
Inclusion of electronic correlations beyond the single-particle
approximation could also be important, however, attempts in this direction have
so far been limited~\cite{delaney,hettler,darancet,thygesen_gw}.

\begin{figure}[b]
\begin{center}
\includegraphics[width=0.9\linewidth]{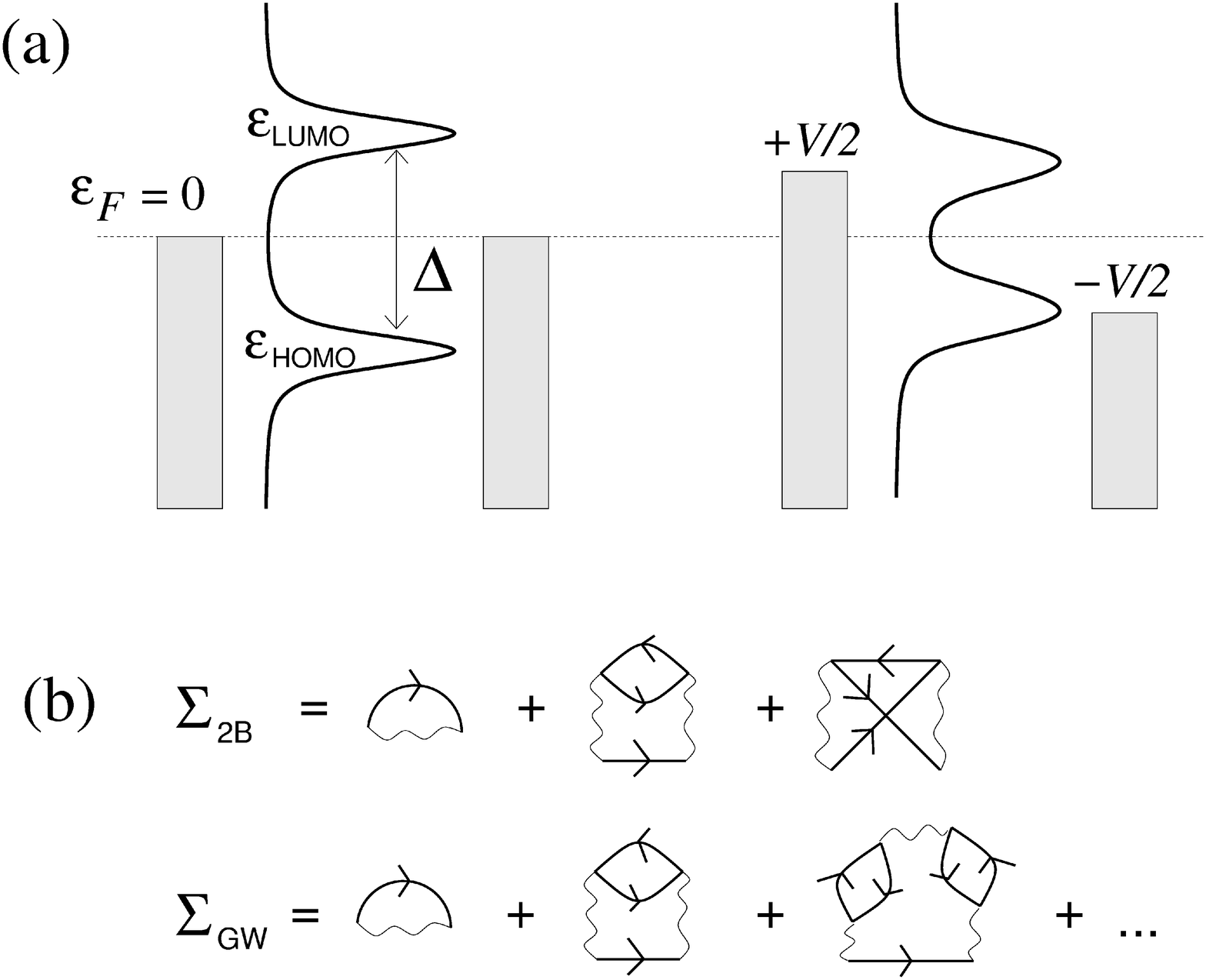}
\end{center}
\caption[system]{\label{fig1} (a) Density of states
  for a molecular junction
  under zero and non-zero bias
  voltage. As indicated the bias changes the line shape of
  the DOS which in turn affects the $dI/dV$. (b) Diagrams for
  the second Born (2B) and $GW$ self-energies. Full lines represent the
Green's function of the
  molecule \emph{with} coupling to leads. Wiggly lines represent the
interaction.}
\end{figure}

While the theory of nano-scale conductance in the
low bias limit and its relation to the ground state electronic
structure has been studied in depth, the finite bias regime has
received much less attention~\cite{delaney,diventra01,bernholc}. The main
reason for this
unbalance is presumably the larger complexity of the latter
problem: linear-response properties can be obtained from the
ground state, e.g. via the Kubo formula~\cite{bokes}, whereas finite-bias
properties
requires the construction of a non-equilibrium steady state which,
unlike the ground state, is not a variational quantity.

In this paper, we analyze the mechanisms governing the shape of molecular
$IV$ characteristics with particular focus on the role of exchange and
correlation. The molecular junction is modeled by two electronic
states representing the HOMO and LUMO levels symmetrically coupled to
leads as sketched in
Fig.~\ref{fig1}(a). It is easy to anticipate that the slope of the
$IV$ curve is largest when a molecular level is aligned with one of
the bias window edges. This will show as peaks in the $dI/dV$ curve. In the
simplest picture the distances between peaks in the $dI/dV$ curve are thus
a direct measure of the distances between the energy levels of the
molecule, and the width of the peaks gives the lifetime of the levels.
This simple picture breaks down because the electronic structure of
the molecule will change in response to the applied bias, and as we
will see, this effect can be surprisingly large even for very simple
systems.

On the basis of the two-level model we identify a simple mechanism
which can lead to the formation of anomalously strong peaks in the
$dI/dV$ curve. The mechanism is driven by the applied bias voltage and
can be viewed as a collapse of the HOMO-LUMO gap by which both levels
move simultaneously into the bias window giving rise to a large
increase in the current. The mechanism is suppressed by SI errors and
should therefore only be partly present in standard DFT transport
calculations. The second result of the paper is that the inclusion of
dynamic correlations, at the levels of the second Born (2B) and $GW$
approximations, can have a strong influence on the $IV$ curve.  This
is mainly due to the enhancement of incoherent QP scattering at
finite bias which leads a to significant broadening of the spectral
features. This effect can also lead
to a suppression of the anomalous $dI/dV$ peaks.

Our model consists of two electronic levels, representing the HOMO and
LUMO states of the molecule, coupled symmetrically to wide-band
leads. The one-particle energies of the levels are $\xi_0$ and
$\xi_0+\Delta_0$, where $\Delta_0$ is the non-interacting
HOMO-LUMO gap. We take the charging energy of the HOMO and LUMO
levels to be the same ($U_{11}$) whereas the charging energy between an electron
in the HOMO and an electron in the LUMO is set to $U_{12}=0.75U_{11}$. As
discussed later our conclusions are not sensitive to this choice
as long as $U_{12}>0.5U_{11}$. We neglect the exchange energy between the HOMO
and LUMO states as it is generally much smaller than the charging
energies~\cite{note}. The Hamiltonian of the molecule is written $\hat
H_{\text{mol}}=\hat H_0+\hat U$, where
\begin{eqnarray}
\hat H_0&=&\sum_{\sigma}\xi_0 \hat
n_{H\sigma}+(\xi_0+\Delta_0)\hat
n_{L\sigma} \\
\hat U&=&\sum_{i=H,L}U_{11}\hat n_{i\uparrow}\hat
n_{i\downarrow}+\sum_{\sigma,\sigma'}U_{12}\hat n_{H\sigma}\hat n_{L\sigma'},
\end{eqnarray}
with $\hat n_{H\sigma}$ and $\hat n_{L\sigma}$ being the number
operator for an electron with spin $\sigma$ in the HOMO and LUMO
states, respectively. The interactions can also be written $\hat
U=\frac{1}{2}\sum_{\sigma,\sigma'}\sum_{i,j\in
\{L,H\}} \tilde U_{i\sigma,j\sigma'}\hat c^{\dagger}_{i\sigma}\hat
c^{\dagger}_{j\sigma'}\hat c_{j\sigma'}\hat c_{i\sigma}$, where
$\tilde U_{i\sigma,j\sigma'}=U_{11}\delta_{ij}(1-\delta_{\sigma
\sigma'})+U_{12}(1-\delta_{ij})$. By working with this
spin-dependent interaction, i.e. using $\tilde U$ to represent the
wiggly lines in the diagrams of Fig.\ref{fig1}(b), SI errors are
automatically avoided to all orders in the interaction since
$\tilde U_{i\sigma,i\sigma}=0$.~\cite{long_gw} This can be illustrated
by noting that both the first- and second-order exchange diagrams
(first and last diagram of $\Sigma_{2B}$) vanish when evaluated using
$\tilde U$.\cite{note2} In the present model these diagrams should
exactly cancel the SI in the Hartree and
second-order bubble diagrams, respectively. By using $\tilde U$ for
the calculation of the $GW$ self-energy we get similar diagrams
included ``for free'' to all orders in the interaction. In particular,
the resulting SI-free $GW$ is exact to second order, i.e. includes the
2B (at least for our model). This is in contrast to standard $GW$
calculations, including previous $GW$ model
calculations~\cite{verdozzi,schindlmayr}, which suffer from higher
order SI errors.\cite{rinke_SIC}

The retarded Green's function describing the molecule in contact with
leads is written
\begin{equation}\label{eq.dyson}
G^{r}(\omega)=[(\omega+i\Gamma) I^{2\times
2}-H_0-\Sigma^{r}_H-\Sigma^{r}(\omega)]^{-1}.
\end{equation}
Spin dependence has been suppressed as we specialize
to the spin unpolarized case, i.e. $G_{\uparrow
  \uparrow}=G_{\downarrow \downarrow}=G$~\cite{note3}. The coupling to leads is
included via the wide-band tunneling rate $\Gamma$. $\Sigma_H$ and
$\Sigma$ denote the Hartree and xc self-energies, respectively. In
this work $\Sigma$ can be either exchange, 2B or (SI-free) $GW$. The
self-energies, $\Sigma_H[G]$ and $\Sigma[G]$, are
calculated self-consistently in conjunction with the Dyson
equation (\ref{eq.dyson}) using the non-equilibrium Keldysh formalism
to account for the difference in chemical potentials. The calculational
procedure is described in detail in Ref.~\onlinecite{long_gw}.

In the special case of symmetric coupling, the particle current can be
written~\cite{meir-wingreen}
\begin{equation}\label{eq.current}
I(V)=\Gamma\int_{\varepsilon_F-V/2}^{\varepsilon_F+V/2} \varrho(V;\omega) \text{d}\omega,
\end{equation}
where $\varrho(\omega)=(i/2\pi)\text{Tr}[G^r(\omega)-G^r(\omega)^{\dagger}]$
is the (non-equilibrium) density of states (DOS). It has peaks at the
position of the QP energy levels, $\varepsilon_i$,
which represent the electron addition/removal energies of the
junction. The width of a peak equals the inverse lifetime of the QP,
$\tau_i^{-1}\approx\Gamma+\text{Im}\Sigma_{ii}^r(\varepsilon_i)$.
Because single-particle approximations are characterized by a real,
frequency independent xc self-energies, the
width of the spectral peaks obtained in Hartree and HF will be given by
$\Gamma$.  From Eq.~(\ref{eq.current}) it is clear that the current
will increase more rapidly when a peak in the DOS enters the bias
window. We stress, however, that $\varrho(\omega)$ depends on $V$ through $\Sigma_H$ and $\Sigma$, and thus the overall shape of
the DOS, in particular the HOMO and LUMO positions, will change with
the bias voltage. Clearly, this change in the DOS determines the shape
of the $dI/dV$ curve.

The following parameters have been used throughout: $\Delta_0=2$, $U_{11}=2$,
$U_{12}=1.5$. By varying the one-particle energy $\xi_0$, we can control the
equilibrium occupation of the molecule, $N_{\text{el}}$. We consider
the case of weak charge transfer to the molecule, i.e. $N_{\text{el}}$ ranges
from 2.0 to 2.1, corresponding to $\varepsilon_F$ lying in the
middle of the gap and slightly below the LUMO, respectively. The
Fermi level is set to zero, and the bias
is applied symmetrically, i.e. $\mu_L=V/2$ and $\mu_R=-V/2$.

\begin{figure}
\begin{center}
\includegraphics[width=0.75\linewidth,angle=270]{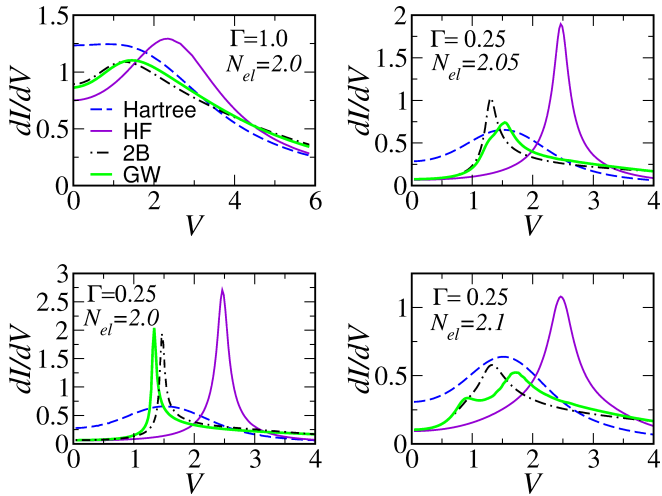}
\end{center}
\caption[system]{\label{fig2} $dI/dV$ curves for different values of
  the tunneling strength $\Gamma$ and occupation of the molecule,
  $N_{\text{el}}$. The curves are calculated using different
  approximations for the xc self-energy.}
\end{figure}

In Fig. \ref{fig2} we show the calculated $dI/dV$ curves (obtained by numerical
differentiation) for different values of
$\Gamma$ and $N_{\text{el}}$. We first notice that the 
$2B$ and $GW$ approximations yield similar results in all the cases indicating
that the higher order terms in the $GW$ self-energy are fairly small. For
$\Gamma=1.0$, all methods yield qualitatively the same
result. For even larger values of $\Gamma$ (not shown), and independently of
$N_{\text{el}}$, the results become even more similar. In this strong
coupling limit, single-particle hybridization effects will dominate over the
interactions.

For all four sets of parameters, the Hartree
approximation severely overestimates the low-bias conductance. This is
a consequence of the SI contained in the Hartree potential which leads
to an underestimation of the (equilibrium) HOMO-LUMO gap. On the other hand the HF,
2B, and $GW$ methods lead to very similar conductances in the low-bias
regime. This indicates that the inclusion of dynamic correlations does not
change the (equilibrium) HOMO-LUMO gap significantly. It is well known
that HF tends to overestimate band gaps, and that the
inclusion of correlations, e.g. at the $GW$ level, has the effect of reducing the HF
gap towards the true value. However, in finite systems where the
electrons are confined in discrete, well separated energy levels,
correlation effects are weak and HF already yields good spectra. In
such cases, the inclusion of correlations is expected to have minor
effects on the equilibrium gap, as observed in the present case. 
We stress, however, that these considerations only apply to weakly
coupled molecular junctions. In fact, for larger values of $\Gamma$ 
the coupling to leads can provide significant screening
of the interactions on the molecule, rendering the $GW$ correction to
the HOMO-LUMO significant~\cite{long_gw}.

Returning to Fig.~\ref{fig2}, we notice that the lower left graph shows an interesting
feature. Namely, the HF, 2B, and $GW$ curves all contain an
anomalously strong conductance peak. Interestingly, the peak height is significantly
larger than 1 which is the maximum conductance for a single
level (the Anderson impurty model). Moreover, the full width at half
maximum (FWHM) of the peak is only
$\sigma_{\text{HF}}=0.27$ and $\sigma_{\text{2B}/GW}=0.12$, respectively,
which is much smaller than the tunneling broadening of $2\Gamma=0.5$. We
note in passing that the peak looses intesity as $N_{\text{el}}$ is
increased, and that the Hartree approximation does not produce the
anomalous peak at all.

\begin{figure}[b]
\begin{center}
\includegraphics[width=0.85\linewidth]{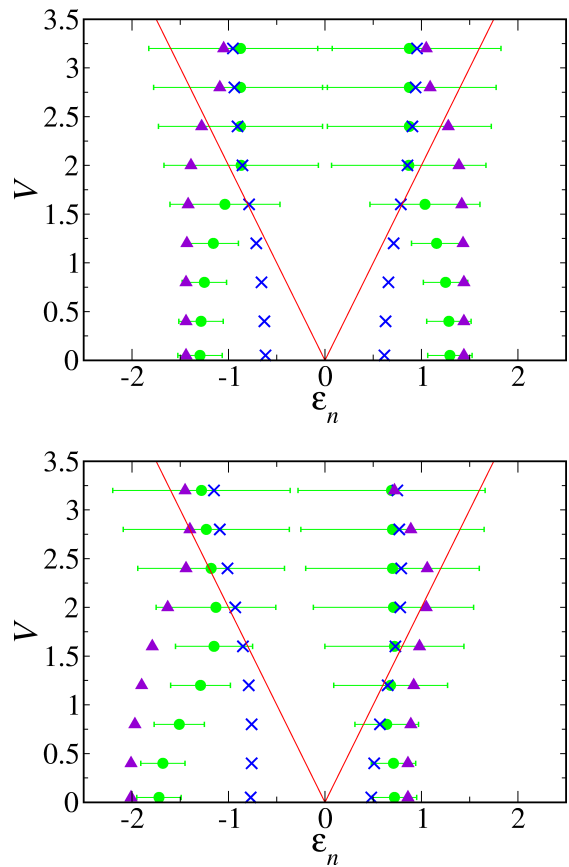}
\end{center}
\caption[system]{\label{fig3} Position of the HOMO and LUMO levels 
as a function of the bias voltage for the Hartree (crosses), HF
(triangles), and $GW$ (circles) approximations. The horizontal lines show the FWHM of
the $GW$ resonances. The FWHM of the Hartree and HF resonances is
$2\Gamma$ independently of $V$. Notice
the differences in the way the levels enter the bias window: The Hartree gap opens while the HF and $GW$ gaps
close. In the upper graph $\Gamma=0.25$,
$N_{\text{el}}=2.0$ (symmetric case). In
the lower graph $\Gamma=0.25$, $N_{\text{el}}=2.1$.}
\end{figure}

To understand the origin of the anomalous peak(s) we plot in Fig.~\ref{fig3}
the evolution of
the HOMO and LUMO positions as a function of the bias
voltage (the 2B result is left out as it is similar to $GW$). Focusing
on the upper panel of the figure ($N_{\text{el}}=2.0$), we notice a
qualitative difference between the Hartree and the SI-free approximations:
While the Hartree gap expands as the levels move into the bias window, 
the HF and $GW$ gaps shrink leading to a dramatic increase in current around $V=2.5$ and
$V=1.3$, respectively. This is clearly the origin of the
anomalous $dI/dV$ peaks. But why do the SI-free gaps collapse?

Let us consider the change in the HOMO and LUMO positions when
$V$ is increased by $2\delta V$. In general this change must be
determined self-consistently, however, a ``first iteration''
estimate yields a change in the HOMO (LUMO) occupations of
$\delta n_H=-\varrho_H(-V/2)\delta V$ ($\delta
n_L=\varrho_L(V/2)\delta V$). At the HF level this leads to 
\begin{eqnarray}\label{eq.ehomo}
\delta \varepsilon_H&=&[-U_{11}
\varrho(-V/2)+2U_{12}\varrho(V/2)]\delta V  \\\label{eq.elumo}
\delta \varepsilon_L&=& [U_{11}\varrho(V/2)-2U_{12}\varrho(-V/2)]\delta V
\end{eqnarray}
where we have used that $\varrho_H(-V/2)\approx \varrho(-V/2)$
and $\varrho_L(V/2)\approx \varrho(V/2)$. The factor 2 in front of
$U_{12}$ includes interactions with both spin channels. In the
symmetric case ($N_{\text{el}}=2.0$) we have
$\varrho(-V/2)=\varrho(V/2)$. Since $U_{11}<2U_{12}$ this means that $\delta \varepsilon_H>0$ and $\delta
\varepsilon_L<0$, i.e. the gap is reduced as $V$ is raised. Moreover it follows that the
gap reduction is largest when $\varrho(\pm V/2)$ is largest, that is, just when the levels cross the bias window. In the
general case ($N_{\text{el}}\neq 2.0$) the direction of the shift depends on
the relative magnitude of the DOS at the two bias window edges: 
a level will follow the edge of the bias window if the other level
does not intersect the bias edge. It will move opposite to the bias,
i.e. into the bias window, if the other
level is close to the bias window edge. This is effect clearly seen in
the lower graph of Fig.~\ref{fig3} (triangles). Thus the gap closing mechanism has the largest impact
on the $dI/dV$ curve when the HOMO and LUMO levels hit the bias window
simultaneously. Moreover, the effect is stronger the larger $U_{12}/U_{11}$, and
the smaller $\Gamma$ (the maximum in the DOS is $\sim 1/\Gamma$).
At the Hartree level, Eqs.~(\ref{eq.ehomo})
and (\ref{eq.elumo}) are modified by replacing $U_{11}$ by $2U_{11}$. This
leads to an effective pinning of the levels to the bias window which tends to open the gap as $V$ is
increased, see Fig.~\ref{fig3} (crosses).

The effect of dynamic correlations can be identified by comparing the HF
result with the $GW$/2B results in Fig.~\ref{fig2}. For
$\Gamma=0.25$ two
qualitative differences are observed: (i) The $GW$/2B conductance
peaks occur at a lower bias voltage than the HF peak. (ii) The
$GW$/2B peaks have long, flat tails on the high-bias side while HF peaks are
more symmetric. Returning to Fig.~\ref{fig3}
we see that part of (i) can be explained from the fact that HF yields
a larger equilibrium gap than $GW$. Indeed, for $V=0$
the HF gap is $\sim 0.3$ larger than the $GW$ gap. However, this effect alone cannot
account for the shift in the conductance peak
from $V\sim 2.5$ in HF to $V\sim 1.5$ in $GW$/2B.

In fact, both (i) and (ii) are consequences of a significant spectral
broadening occurring at finite bias in the $GW$/2B calculations. The
broadening, indicated by horizontal lines in Fig. \ref{fig3}, is due to QP scattering. 
According to
Fermi-liquid theory, QP scattering at the Fermi level
is strongly suppressed in the ground state,
i.e. $\text{Im}\Sigma_{ii}(\varepsilon_F)=0$ for $V=0$. However, as the
bias is raised the phase space available for QP scattering is enlarged and
$\text{Im}\Sigma$ increases accordingly. As a result of the additional
level broadening, $\varrho(\pm V/2)$ increases more rapidly as a
function of $V$. Since this is exactly the driving force behind the gap
closing mechanism, the $dI/dV$ peak occurs earlier in the $GW$ than
the HF calculation. Clearly, the long tails seen in the $dI/dV$ of the
$GW$/2B calculations are also a result of the spectral broadening due to
QP scattering.

In summary, we have investigated the mechanisms governing the shape of
the $IV$ characteristic of a generic molecular junction. We identified a
simple gap-closing mechanism which can
lead to anomalously strong peaks in the $dI/dV$. The mechanism is
suppressed by SI errors and is therefore not likely to be correctly
described in standard DFT calculations.  This shows that the use of SI
corrected xc-functionals, recently shown to be important for the
linear-response conductance, could be equally important under finite
bias conditions. Finally, we found that the strong enhancement of QP
scattering as function of bias voltage leads to significant smearing
of spectral features, which in turn has a large impact on the $dI/dV$ curve.

The author thanks Angel Rubio for useful discussions and Duncan
Mowbray for critical reading of the manuscript. This work was
supported by The Lundbeck Foundation's Center for
Atomic-scale
Materials Design (CAMD) and the Danish Center for
Scientific Computing through grant No. HDW-1103-06.



\begin{thebibliography}{30}
\bibitem{note}
In the case of benzene we have calculated the relevant charging
energies and find
$U_{\text{HOMO}}\approx 8.5$eV, $U_{\text{LUMO}}\approx 7.7$eV, and
$V\approx 7.0$eV. The exchange energy
between HOMO and LUMO states is $V_x=1.1$eV.
\bibitem{note2}
This follows by noting that the Green's function, $G$, is diagonal in
the $(i\sigma)$ indices (there is no coupling between the levels)
while the diagonal elements of $\tilde V$ are zero.
\bibitem{note3}
For the considered
parameter values we have found no
spin-polarized solutions for the Hartree, 2B and $GW$ approximations. For
$\Gamma=0.2$ a spin-polarized HF
solution exists when one of the levels
come sufficiently close to $\mu_R$ or $\mu_L$.
\end{thebibliography}
\end{document}